\documentclass[11pt]{article}
\usepackage{amssymb,latexsym, amsmath}
\usepackage{graphicx}
\makeatletter
\@addtoreset{equation}{section}

\makeatother

\newtheorem{theorem}{Theorem}[section]
      
\newtheorem{corollary}[theorem]{Corollary}

\newtheorem{rem}[theorem]{Remark}

\newtheorem{exa}[theorem]{Example}

\newtheorem{ques}[theorem]{Question}

\def\mathbb#1{\mathbf{ #1}}
\def\mathfrak#1{\mathbf{ #1}}
\begin{document}

\title{\vbox to 0pt 
{\vskip -2cm \rlap{\hbox to .5\textwidth 
{\rm{
{\small  SUBMITTED TO: }
{\small\it Journal of Physics A}
} } 
} }
\vspace{-6mm}
{\it Variational Principles for
Lagrangian Averaged Fluid Dynamics}
}
\author{
Darryl D. Holm\footnote{ email: dholm@lanl.gov}
\\Theoretical Division and Center for Nonlinear Studies
\\Los Alamos National
Laboratory, MS B284
\\ Los Alamos, NM 87545
}

\date{{\small SUBMITTED March 22, 2001, TO:
\\
\it Journal of Physics A}}

\maketitle

\large

\begin{abstract}
The Lagrangian average (LA) of the ideal fluid equations preserves
their transport structure. This transport structure is responsible
for the Kelvin circulation theorem of the LA flow and, hence, for its
convection of potential vorticity and its conservation of helicity.

Lagrangian averaging also preserves the Euler-Poincar\'e (EP)
variational framework that implies the LA fluid equations. This is
expressed in the Lagrangian-averaged Euler-Poincar\'e (LAEP) theorem
proven here and illustrated for the Lagrangian average Euler (LAE)
equations.



\end{abstract} 


\section{Introduction}
\label{Intro-sec}


\paragraph{Decomposition of multiscale problems {and} scale-up}
In turbulence, in climate modeling and in other multiscale
fluids problems, a major challenge is ``scale-up.'' This is the
challenge of deriving models that correctly capture the mean, or
large scale flow -- including the influence on it of the rapid,
or small scale dynamics. 

In classical mechanics this sort of problem has been approached
by choosing a proper ``slow + fast'' decomposition and deriving
evolution equations for the slow mean quantities by using, say,
the standard method of averages.  For nondissipative systems in
classical mechanics that arise from Hamilton's variational principle,
the method of averages may extend to the averaged Lagrangian method,
under certain conditions. 


\paragraph{Eulerian vs Lagrangian means}
In meteorology and oceanography, the averaging approach has a
venerable history and many facets. Often this averaging is
applied in the geosciences in combination with additional
approximations involving force balances (for example,
geostrophic and hydrostatic balances). It is also sometimes
discussed as an initialization procedure that seeks a nearby
invariant ``slow manifold.''  Moreover, in meteorology and
oceanography, the averaging may be performed in either the
Eulerian, or the Lagrangian description. The relation between
averaged quantities in the Eulerian and Lagrangian descriptions
is one of the classical problems of fluid dynamics.
 

\paragraph{Generalized Lagrangian mean (GLM)}
The GLM equations of Andrews {and} McIntyre [1978a] systematize
the approach to Lagrangian fluid modeling by introducing a slow + fast
decomposition of the Lagrangian particle trajectory in general form.
In these equations, the Lagrangian mean of a fluid quantity evaluated
at the mean particle position is related to its Eulerian mean,
evaluated at the displaced fluctuating position. The GLM equations are
expressed directly in the Eulerian representation. The Lagrangian mean
has the advantage of preserving the fundamental transport structure of
fluid dynamics. In particular, the Lagrangian mean commutes with the
scalar advection operator and it preserves the Kelvin circulation
property of the fluid motion equation.


\paragraph{Compatibility of averaging and reduction of
Lagrangians for mechanics on Lie groups} 
In making slow + fast decompositions and constructing averaged
Lagrangians for fluid dynamics, care must generally be taken to
see that the averaging and reduction procedures do not interfere
with each other. Reduction in the fluid context refers to symmetry
reduction of the action principle by the subgroup of the
diffeomorphisms that takes the Lagrangian representation to the
Eulerian representation of the flow field. The theory for this
yields the  Euler-Poincar\'e (EP) equations, see 
Holm, Marsden {and} Ratiu [1998a,b] and Marsden {and} Ratiu [1999].


\paragraph{Lagrangian averaged Euler-Poincar\'e (LAEP) equations}
The compatibility requirement between averaging and reduction is
handled automatically in the Lagrangian averaging (LA) approach. The
Lagrangian mean of the action principle for fluids does not interfere
with its reduction to the Eulerian representation, since the averaging
process is performed at {\it fixed Lagrangian coordinate}. Thus, the
process of taking the Lagrangian mean is compatible with reduction by
the particle-relabeling group of symmetries for Eulerian fluid
dynamics.

In this paper, we perform this reduction of the action principle and
thereby place the LA fluid equations such as GLM theory into the EP
framework. In doing this, we demonstrate the variational reduction
property of the Lagrangian mean. This is encapsulated in the LAEP
Theorem proven here:

\begin{theorem}[Lagrangian Averaged Euler-Poincar\'e Theorem]
The Lagrangian averaging process preserves the variational structure
of the Euler-Poincar\'e framework.
\end{theorem}

According to this theorem, the Lagrangian mean's preservation
of the fundamental transport structure of fluid dynamics also
extends to preserving the EP variational structure of these
equations. This preservation of structure may be {\it visualized as a
cube} in Figure 1. As we shall explain, the LAEP theorem produces a cube
consisting of four equivalence relations on each of its left and right
faces, and four commuting diagrams (one on each of its four remaining
faces).


%
\begin{figure}\vspace{-14mm}
\begin{center}
\includegraphics[scale=0.4,angle=0]{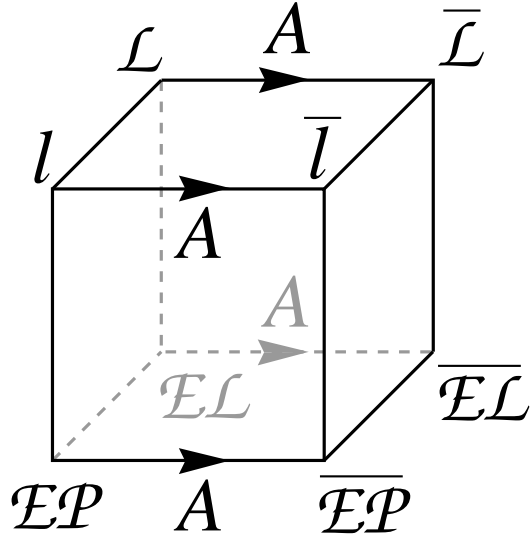}
\end{center}\vspace{-12mm}
\caption{\label{cp}{\footnotesize
The Averaged EP theorem produces a cube consisting of four equivalence
relations on each of its left and right faces, and four commuting
diagrams (one on each of its four remaining faces).}}
\end{figure}
%


\paragraph{Euler-Lagrange-Poincar\'e (ELP) cube} The front and back
faces of the ELP cube live in the Eulerian (spatial) and Lagrangian
(material) pictures of fluid dynamics, respectively. The top face
contains four variational principles at its corners and the bottom face
contains their corresponding equations of motion. The horizontal edges
represent Lagrangian averaging and are directed from the left to the
right. The left face contains the four equivalence relations of the
Euler-Poincar\'e Theorem on its corners and the right face contains the
corresponding averaged equivalence relations. Thus, the left and right
faces of the ELP cube are equivalence relations, and its front, back,
top and bottom faces are commuting diagrams.

The back face of the ELP cube displays the LA 
preservation of variational structure in the Lagrangian fluid
picture. Hamilton's principle with $L$ yields the Euler-Lagrange
equations $EL$ in this picture, and Lagrangian averaging $A$ preserves
this relation. Namely, Hamilton's principle with the averaged
Lagrangian $\bar{L}$ yields the averaged Euler-Lagrange equations
$\overline{EL}$. 

This pair of Hamilton's principles and Euler-Lagrange equations
has its counterpart in the Eulerian picture of fluid dynamics Ð
on the front face of the ELP cube -- whose variational relations are
also exactly preserved by the LA process. 

The bottom front edge of the cube represents the GLM equations of
Andrews {and} McIntyre [1978a]. Thus, the GLM equations represent a
foundational result for the present theory.

The six faces of the ELP cube represent six
interlocking equivalence relations and commutative diagrams that
enable modeling and Lagrangian averaging to be performed equivalently
either at the level of the equations, as in  Andrews {and} McIntyre
[1978a], or at the level of Hamilton's principle. At the level of
Hamilton's principle, powerful theorems from other mean field theories
are available. An example is Noether's theorem, which relates
symmetries of Hamilton's principle to conservation laws of the
equations of motion. Fluid conservation laws include mass, momentum and
energy, as well as local conservation of potential vorticity. The
latter yields the Casimirs of the corresponding Lie-Poisson Hamiltonian
formulation of ideal fluid dynamics and is due to the symmetry of
relabeling diffeomorphisms admitted by Hamilton's principle for fluid
dynamics, see Arnold [1966] and Holm, Marsden, Ratiu {and} Weinstein
[1985]. In certain cases, the fluid vorticity winding number (called
helicity -- a topological quantity) is also conserved. Lagrangian
averaging preserves all of these conservation laws.%
\footnote{We note that the conserved mean topological quantity resulting
after Lagrangian averaging is the helicity of the mean fluid vorticity,
not the mean of the original helicity.}
Thus, the LA Hamilton's principle yields the LA fluid equations in
either the Lagrangian, or the Eulerian fluid picture, and one may
transform interchangeably along the edges of the cube in search of
physical and mathematical insight.

\begin{rem}[Eulerian mean]
Of course, the preservation of variational structure resulting in the
interlocking commuting relationships and conservation laws of the LAEP
Theorem is not possible with the Eulerian mean, because the Eulerian
mean does {\it not} preserve the transport structure of fluid mechanics.
\end{rem}

\begin{rem}[Balanced approaches]
The LAEP Theorem puts the approach using averaged Hamilton's
principles and the method of Lagrangian averaged equations onto
equal footing. This is quite a bonus for both approaches to modeling
fluids. According to the LAEP Theorem, the averaged Hamilton's
principle produces dynamics that is guaranteed to be verified
directly by averaging the original equations, and the Lagrangian
averaged equations inherit the conservation laws that are available
from the symmetries of Hamilton's principle.
\end{rem}


\paragraph{Outline of the paper}

In section \ref{EP-sec}, we begin by briefly reviewing the
mathematical setting of the EP theorem from Holm, Marsden {and} Ratiu
[1998a,b]. We state the EP theorem and
discuss a few of its implications for continuum mechanics in vector
notation. We also sketch its proof, in preparation for proving
the corresponding results for the Lagrangian Averaged
Euler-Poincar\'e (LAEP) theorem presented in section \ref{LAEP-sec}.
Finally, in section \ref{Applic-incomp-sec}, we illustrate the LAEP
theorem by applying it to incompressible ideal fluids. We also mention
recent progress toward closure of these equations as models of fluid
turbulence.



\section{The Euler-Poincar\'e theorem
for fluids with advected properties}
\label{EP-sec}


\subsection{Mathematical setting and statement of the EP theorem}


The assumptions of the Euler-Poincar\'e
theorem from Holm, Marsden {and} Ratiu [1998a] are briefly
listed below.
\begin{description}

\item$\quad\bullet$ There is a {\it right\/} representation of Lie
group $G$ on the vector space $V$ and $G$ acts in the natural way on
the {\it right\/} on $TG \times V^\ast$: $(v_g, a)h = (v_gh, ah)$.

\item$\quad\bullet$ Assume that the function $ L : T G \times V ^\ast
\rightarrow \mathbb{R}$ is right $G$--invariant.

\item$\quad\bullet$ In particular, if $a_0 \in V^\ast$, define the
Lagrangian $L_{a_0} : TG \rightarrow \mathbb{R}$ by
$L_{a_0}(v_g) = L(v_g, a_0)$. Then $L_{a_0}$ is right
invariant under the lift to $TG$ of the right action of
$G_{a_0}$ on $G$, where $G_{a_0}$ is the isotropy group of $a_0$.

\item$\quad\bullet$  Right $G$--invariance of $L$ permits one to define
$\ell: {\mathfrak{g}} \times V^\ast \rightarrow \mathbb{R}$ by
\begin{equation}
\ell(v_gg^{-1}, a_0g^{-1}) = L(v_g, a_0).
\nonumber
\end{equation}
Conversely,  this relation defines for any
$\ell: {\mathfrak{g}} \times V^\ast \rightarrow
\mathbb{R} $ a right $G$--invariant function
$ L : T G \times V ^\ast
\rightarrow \mathbb{R} $.

\item$\quad\bullet$ For a curve $g(t) \in G, $ let
\begin{equation}
u(t) \equiv \dot{g}(t) g(t)^{-1}\in TG/G\cong \mathfrak{g}
\nonumber
\end{equation}
and define the curve $a(t)$ as the unique solution of the linear
differential equation with time dependent coefficients 
\begin{equation}\label{advection-rel}
\dot a(t) = -a(t)u(t)
\end{equation}
where the action of $u\in\mathfrak{g}$ on the initial condition $a(0)
= a_0\in V^*$ is denoted by concatenation from the right. The
solution of (\ref{advection-rel}) can be written as the {\bf advective
transport relation},
\begin{equation}
a(t)
= a_0g(t)^{-1}
\,.\nonumber
\end{equation}

\end{description}

\begin{theorem} [EP Theorem]\label{rarl}

The following are equivalent:
\begin{enumerate}

\item [{\bf i} ] Hamilton's variational principle
\begin{equation} \label{hamiltonprincipleright1}
\delta \int _{t_1} ^{t_2} L_{a_0}(g(t), \dot{g} (t)) dt = 0
\end{equation}
holds, for variations $\delta g(t)$
of $ g (t) $ vanishing at the endpoints.

\item [{\bf ii}  ] $g(t)$ satisfies the Euler--Lagrange
equations for $L_{a_0}$ on $G$.

\item [{\bf iii} ]  The constrained variational principle
\begin{equation} \label{variationalprincipleright1}
\delta \int _{t_1} ^{t_2}  \ell\,(u(t), a(t)) dt = 0
\end{equation}
holds on $\mathfrak{g} \times V ^\ast $, using variations of the form
\begin{equation} \label{variationsright1}
\delta u = \frac{ \partial \eta}{\partial t } + {\rm ad}_u\,\eta,
\quad
\delta a =  -a\,\eta ,
\end{equation}
where $\eta(t) \in \mathfrak{g}$ vanishes at the
endpoints.

\item [{\bf iv}] The Euler--Poincar\'{e} equations hold on
$\mathfrak{g} \times V^\ast$
\begin{equation} \label{eulerpoincareright1}
\frac{ \partial}{\partial t} \frac{\delta \ell}{\delta u} 
= 
-
 {\rm ad}_{u}^{\ast} \frac{ \delta \ell }{ \delta u}
+ 
\frac{\delta \ell}{\delta a} \diamond a.
\end{equation}
\end{enumerate}
\end{theorem}


\subsection{Discussion of the EP equations in vector notation}


When mass is the {\it only} advected quantity, the EP motion equation
(\ref{eulerpoincareright1}) and the advection relation
(\ref{advection-rel}) for mass conservation are written as 
\begin{equation}
\Big(
\frac{ \partial}{\partial t}
+
{\rm ad}^*_u
\Big)
\frac{ \delta \ell}{ \delta {u}}
-
\frac{ \delta \ell}{ \delta {D}}\diamond D
=
0
\,,\quad\hbox{and}\quad
\frac{ \partial D}{\partial t}
=
-\,\pounds_u D
\,.
\nonumber
\end{equation}
Here $\pounds_u$ denotes the Lie derivative with respect to velocity
${u}$, and the operations  ad${^*}$ and
$\diamond$ are defined using the $L_2$ pairing 
$\langle{f,g}\rangle=\int fg\, d^3x$. The ad$^*$ operation is defined
as (minus) the $L_2$ dual of the Lie algebra operation, ad, for vector
fields, namely 
\begin{equation}
-\,
\Big\langle
{\rm ad}^*_u\, \frac{ \delta \ell}{ \delta {u}}, \eta
\Big\rangle
=
\Big\langle
\frac{ \delta \ell}{ \delta {u}}, {\rm ad}_u\,\eta
\Big\rangle
\,.\nonumber
\end{equation}
In vector notation ad$_u\,\eta$ is expressed as
\begin{equation}
{\rm ad}_u\,\eta 
= 
\eta u - u\eta 
=
-\,[u,\eta\,]
=
-\,{\rm ad}_\eta\, u
= 
u\cdot\nabla\eta - \eta\cdot\nabla u
\,.\nonumber
\end{equation}
The diamond operation $\diamond$ is defined as (minus) the $L_2$ dual
of the Lie derivative, namely, 
\begin{equation}
-\,\Big\langle
\frac{ \delta \ell}{ \delta a}\diamond a,\eta
\Big\rangle
=
\Big\langle
\frac{ \delta \ell}{ \delta a},
\pounds_\eta a
\Big\rangle
\,.
\nonumber
\end{equation}
Here $a$ and $\delta \ell/\delta a$ are dual tensors
and $(\delta \ell/\delta a)\diamond a$ is a one-form density (dual to
vector fields under $L_2$ pairing). In vector notation the diamond
operation $\diamond$ for the example of the density $D$ becomes
\begin{equation}
-\,\Big\langle
\frac{ \delta \ell}{ \delta {D}}\diamond D,\eta
\Big\rangle
=
\Big\langle
\frac{ \delta \ell}{ \delta {D}},
\pounds_\eta D
\Big\rangle
=
\Big\langle
\frac{ \delta \ell}{ \delta {D}},
{\rm div}(D\eta) 
\Big\rangle
=
-\,
\Big\langle
D\nabla\frac{ \delta \ell}{ \delta {D}},
\eta
\Big\rangle
\,.
\nonumber
\end{equation}
Thus, the EP motion equation (\ref{eulerpoincareright1}) may be
written in Cartesian components as
\begin{equation}
\frac{ \partial}{\partial t}
\frac{ \delta \ell}{ \delta {u}^i}
+ 
\underbrace{\,
\frac{ \partial}{\partial {x}^j}
\Big(
\frac{ \delta \ell}{ \delta {u}^i}{u}^j
\Big)
+
\frac{ \delta \ell}{  \delta u^j}
\frac{ \partial {u}^j}{\partial {x}^i}
}
_{\equiv\,\big({\rm ad}^*_u
\,
\frac{ \delta \ell}{ \delta {u}}\big)_i
\,}
-
\underbrace{\,
D
\frac{ \partial }{ \partial {x}^i}
\Big(
\frac{ \delta \ell}{ \delta {D}}
\Big)
\,}
_{\equiv\,\big(\frac{ \delta \ell}{ \delta {D}}\,\diamond \, D
\big)_i
}
=0
\,,
\nonumber
\end{equation}
and the advection relation for the mass density $D\in V^*$ satisfies,
\begin{equation}
\Big(
\frac{ \partial}{\partial t}
+
\pounds_u
\Big)
\Big(D\,d^3x\Big)
=
0
\,,\quad\hbox{or}\quad
\frac{ \partial D}{\partial t}
=
-\,{\rm div}(D\mathbf{u})
\,.
\nonumber
\end{equation}
%

\noindent
{\bf Remarks.}
\begin{description}

\item$\quad\bullet$
In passing from coordinate-free forms to their component expressions,
we shall write tensors in a Cartesian basis. For example, we shall
include the volume form in the mass density by denoting it as
$Dd^3x$.

\item$\quad\bullet$
The EP motion equation and mass advection equation may also be
written equivalently using Lie derivative notation as 
\begin{equation}
\Big(
\frac{ \partial}{\partial t}
+
\pounds_u
\Big)
\frac{ \delta \ell}{ \delta {u}}
-
\frac{ \delta \ell}{ \delta {D}}\diamond D
=
0
\,,\quad\hbox{and}\quad
\Big(
\frac{ \partial}{\partial t}
+
\pounds_u
\Big)
D
=
0
\,.
\nonumber
\end{equation}
The equivalence here of $\pounds_u$ and ad$^*_u$ arises
because $\delta\ell/\delta u$ is a one-form density and the equality 
ad$^*_u\mu=\pounds_u\mu$ holds for any one-form density $\mu$.

\item$\quad\bullet$
In the Lie derivative notation, one proves the {\bf Kelvin-Noether
circulation theorem} immediately as a corollary, by
\begin{equation}
\frac{d}{dt}
\oint_{c(u)}
\frac{1}{D}\
\frac{ \delta \ell}{ \delta {u}}
=
\oint_{c(u)}\!\!
\Big(
\frac{ \partial}{\partial t}
+
\pounds_u
\Big)
\frac{1}{D}\
\frac{ \delta \ell}{ \delta {u}}
=
\oint_{c(u)}
\frac{1}{D}\
\frac{ \delta \ell}{ \delta {D}}\diamond D
\,,
\nonumber
\end{equation}
for any closed curve $c(u)$ that moves with the fluid.
In vector notation, this is seen as
\begin{equation}\label{Kelvin-circ-eqn}
\frac{d}{dt}
\oint_{c(u)}
\frac{1}{D}\
\frac{ \delta \ell}{ \delta \mathbf{u}}
\cdot
d\mathbf{x}
=
\oint_{c(u)}
\nabla\frac{ \delta \ell}{ \delta {D}}
\cdot
d\mathbf{x}
=
0\,.
\end{equation}

\item$\quad\bullet$
{\bf Helicity conservation.} In Lie derivative notation, one may
rewrite the Kelvin circulation theorem as 
\begin{eqnarray*}
\Big(
\frac{ \partial}{\partial t}
+
\pounds_u
\Big) v
+
dp
=
0
\,,
\end{eqnarray*}
where
\begin{eqnarray*}
v
=
\mathbf{v}\cdot d\mathbf{x}
=
\frac{1}{D}\
\frac{ \delta \ell}{ \delta \mathbf{u}}
\cdot
d\mathbf{x}
\,,\quad\hbox{and}\quad
p = -\, \frac{ \delta \ell}{ \delta {D}}
\,.
\end{eqnarray*}
Therefore,
\begin{eqnarray*}
\Big(
\frac{ \partial}{\partial t}
+
\pounds_u
\Big) (v\wedge dv)
=
-d\,(p\, dv)
\,,
\end{eqnarray*}
where $\wedge$ is the exterior product of differential forms. In
vector notation, this is the helicity equation,
\begin{equation}\label{Helicity-eqn}
\frac{ \partial}{\partial t}
(\mathbf{v}\cdot {\rm curl}\,\mathbf{v})
+
{\rm div}\,\big(
\mathbf{u}\,(\mathbf{v}\cdot {\rm curl}\,\mathbf{v})
+
p\ 
{\rm curl}\,\mathbf{v}
\big)
=
0
\,.
\end{equation}
Consequently, for homogeneous boundary conditions one finds conservation
of helicity 
$\Lambda=\int v\wedge dv 
= 
\int \mathbf{v}\cdot {\rm curl}\,\mathbf{v}\, d^3x$. The helicity
$\Lambda$ is a topological quantity that measures the linkage number of
the lines of curl$\,\mathbf{v}$, the fluid vortex lines in this case.

\end{description}


\subsection{Proof of the EP Theorem } 


The equivalence of {\bf i} and {\bf ii} holds for any configuration
manifold and so, in particular, it holds in this case.

The following string of equalities
shows that {\bf iii} is equivalent to {\bf iv}.
\begin{eqnarray}\label{continuumEPderivation}
0 
&=&
\delta \int_{t_1}^{t_2}\!\!
   \ell(u, a)\, dt 
=
\int_{t_1}^{t_2}
\Big\langle
\frac{\delta \ell}{\delta u}\,,\,\delta u 
\Big\rangle
+
\Big\langle
\frac{\delta \ell}{\delta a}\,,\,\delta a 
\Big\rangle
\,dt 
\nonumber \\
  &=& 
 \int_{t_1}^{t_2}
  \left\langle\frac{\delta \ell}{\delta u}
  \,,\, 
  \frac{\partial\eta}{\partial t}
    +{\rm ad}_u\,\eta
 \right\rangle
    -
\left\langle
\frac{\delta \ell}{\delta a}\,,\, \pounds_\eta\, a
\right\rangle
\,dt
\nonumber \\
   &=&-\int_{t_1}^{t_2}
 \left\langle
 \frac{\partial}{\partial t}\frac{\delta \ell}{\delta u}
   + {\rm ad}^*_u\frac {\delta \ell}{\delta u}
   -\frac{\delta \ell}{\delta a} \diamond a 
\,,\,
\eta
\right\rangle
dt
\,.
\end{eqnarray}

Finally we show that {\bf i} and {\bf iii} are equivalent.
First note that the $G$--invariance of $L:TG
\times V^\ast \rightarrow \mathbb{R}$ and the definition of 
$a(t) = a_0g(t)^{-1}$ imply that the
integrands in (\ref{hamiltonprincipleright1}) and
(\ref{variationalprincipleright1}) are equal. Moreover, all variations
$\delta g(t) \in TG$ of $g(t)$ with fixed endpoints induce and are
induced by variations $\delta u(t) \in
\mathfrak{g}$ of $u(t)$ of the form 
$\delta u = \partial\eta /\partial t + {\rm ad}_u\,\eta $ with 
$\eta(t) \in \mathfrak{g}$
vanishing at the endpoints. The relation between $\delta g(t)$ and
$\eta(t)$ is given by $\eta(t) = \delta g(t)g(t)^{-1}$.
This is the proof first given in Holm, Marsden {and} Ratiu [1998a].
\par\hfill QED\smallskip


\section{Lagrangian averaged Euler-Poincar\'e theory} 
\label{LAEP-sec}


We shall place the GLM (Generalized Lagrangian Mean) theory of
Andrews {and} McIntyre [1978a] into the Euler-Poincar\'e framework
discussed in the previous section.


\subsection{GLM theory from a geometric viewpoint} 


The GLM theory of Andrews {and} McIntyre [1978a] begins by assuming the
Lagrange-to-Euler map {\it factorizes} as a product of
diffeomorphisms,
\begin{eqnarray*}
g(t)=\Xi\,(t)\cdot \tilde{g}(t)
\,.
\end{eqnarray*}
Moreover, the first factor $\tilde{g}(t)$ arises from an averaging
process,
\begin{eqnarray*}
\bar{g}(t)
=
\overline{\Xi\,(t)\cdot \tilde{g}(t)}
=
\tilde{g}(t)
\,,
\end{eqnarray*}
that satisfies the {\bf projection property}, so that 
$\bar{\tilde{g}}(t)=\tilde{g}(t)$. 
Thus, a fluid parcel labeled by $\mathbf{x}_0$ has 
current position,
\begin{eqnarray*}
\mathbf{x}^\xi(\mathbf{x}_0,t)
\equiv
\Xi\,(t)\cdot (\tilde{g}(t)\cdot\mathbf{x}_0)
=
\Xi\,(\mathbf{x}(\mathbf{x}_0,t),t)
\quad\hbox{(current position)}
\,,
\end{eqnarray*}
and it has mean position, 
\begin{eqnarray*}
\mathbf{x}(\mathbf{x}_0,t)
=
\tilde{g}(t)\cdot\mathbf{x}_0
\quad\hbox{(mean position)}
\,.
\end{eqnarray*}

\begin{rem}
Thus, GLM theory first averages the action of the diffeomorphism
group, while holding fixed the material objects on which the group
acts. Then it restores the original action of the group by assuming
that $g(t)\cdot \tilde{g}^{-1}(t)=\Xi\,(t)$ is also a diffeomorphism.
This is illustrated in Figure 2.
\end{rem}


%
\begin{figure}\vspace{-10mm}
\begin{center}
\includegraphics[scale=0.6,angle=0]{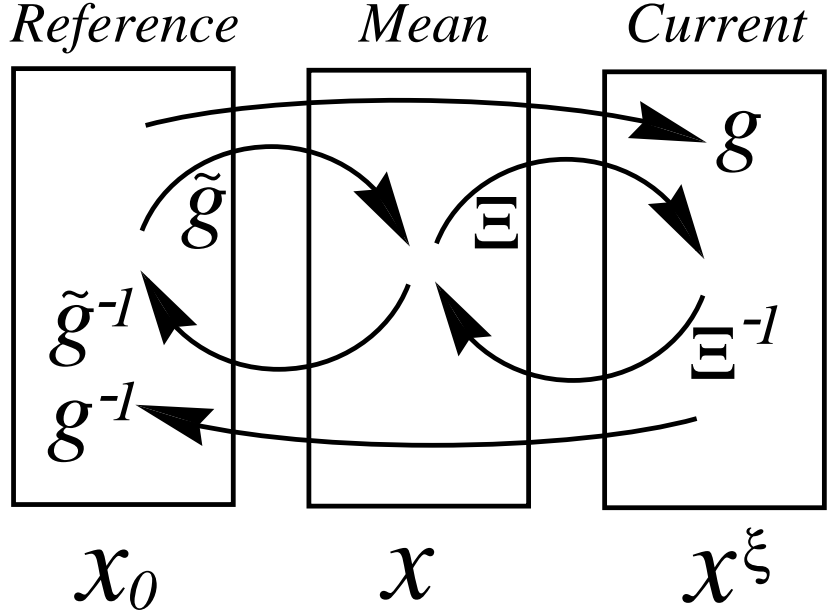}
\end{center}\vspace{-12mm}
\caption{\label{cp}{\footnotesize
GLM theory factorizes the Lagrange to Euler map at a given time by
first mapping the reference configuration to the mean position, then
mapping that to the current position.}}
\end{figure}
%



%
\begin{figure}\vspace{-13mm}
\begin{center}
\includegraphics[scale=0.5,angle=0]{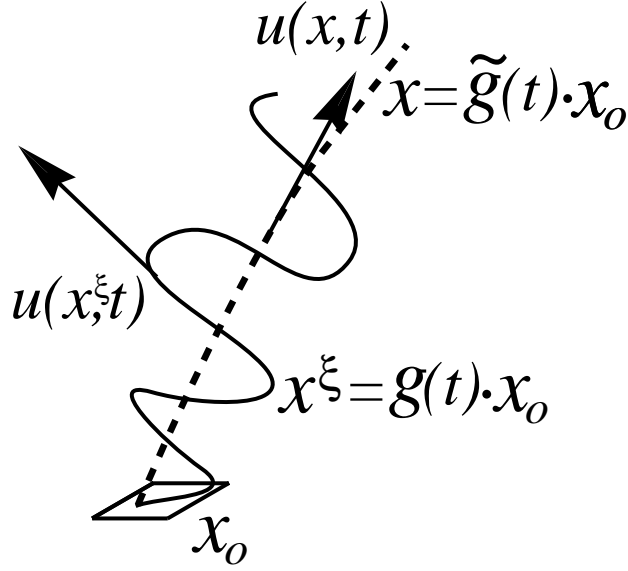}
\end{center}\vspace{-12mm}
\caption{\label{cp}{\footnotesize
The GLM velocities $\mathbf{u}(\mathbf{x}^\xi,t)$ and
$\bar\mathbf{u}^L(\mathbf{x},t)$ are tangent to the current and mean
trajectories, $\mathbf{x}^\xi$ and $\mathbf{x}$, respectively.}}
\end{figure}
%


The composition of maps $g(t)=\Xi\,(t)\cdot \tilde{g}(t)$ yields via
the chain rule the following {\bf velocity relation},
\begin{equation}\label{vee-rel}
\dot{g}(t)\cdot \mathbf{x}_0
=
\dot{\Xi}\,(t)\cdot\mathbf{x}
+
T\Xi\cdot\!(\dot{\tilde{g}}(t)\cdot \mathbf{x}_0)
\,.
\end{equation}
By invertibility, 
$\mathbf{x}_0
=
g^{-1}(t)\cdot\mathbf{x}^\xi
=
\tilde{g}^{-1}(t)\cdot\mathbf{x}
$. Consequently, one may define the fluid parcel velocity at the current
position in terms of a vector field evaluated at the mean position as,
\begin{eqnarray*}
\mathbf{u}(\mathbf{x}^\xi,t)
=
\dot{g}\cdot g^{-1}(t)\cdot\mathbf{x}^\xi
=
\dot{g}\cdot \tilde{g}^{-1}(t)\cdot\mathbf{x}
=
\mathbf{u}^\xi(\mathbf{x},t)
\,.
\end{eqnarray*}
Hence, by using the velocity relation (\ref{vee-rel}) one finds, 
\begin{eqnarray}\label{u-xi-def}
\mathbf{u}^\xi(\mathbf{x},t)
&=&
\dot{\Xi}\,(t)\cdot \mathbf{x}
+
T\Xi\cdot\!
\big(\,\dot{\tilde{g}}\tilde{g}^{-1}(t)\cdot \mathbf{x}\,\big)
\nonumber\\
&\equiv&
\frac{\partial \Xi}{\partial t} \,(\mathbf{x},t)
+
\frac{\partial \Xi}{\partial \mathbf{x}}
\cdot
\bar{\mathbf{u}}^L(\mathbf{x},t)
\,.
\end{eqnarray}
Here the {\bf Lagrangian mean velocity} $\bar{\mathbf{u}}^L$ is defined
as
\begin{equation}\label{LA-vee}
\bar{\mathbf{u}}^L(\mathbf{x},t)
\equiv
\overline{\mathbf{u}^\xi(\mathbf{x},t)}
=
\overline{\dot{g}\tilde{g}^{-1}(t) }\cdot \mathbf{x}
=
\dot{\tilde{g}}(t)\tilde{g}(t)^{-1}
\cdot
\mathbf{x}
=
\dot{\tilde{g}}(t)\cdot \mathbf{x}_0
\,.
\end{equation}
In the third equality we used the projection property of the averaging
process and found
$\overline{\dot{g}}=\dot{\overline{g}}=\dot{\tilde{g}}$ from equation
(\ref{vee-rel}), so that
\begin{eqnarray*}
\bar{\mathbf{u}}^L(\mathbf{x},t)
=
\dot{\tilde{g}}(t)\tilde{g}(t)^{-1}
\cdot
\mathbf{x}
\equiv
\tilde{\mathbf{u}}(\mathbf{x},t)
\,.
\end{eqnarray*}
Thus, the Lagrangian mean velocity vector satisfies
$\bar{\mathbf{u}}^L=\tilde{\mathbf{u}}$, so $\bar{\mathbf{u}}^L$ is
tangent to the mean motion associated with $\tilde{g}(t)$. Hence, one
may write equation (\ref{u-xi-def}) in terms of the  {\bf mean material
time derivative}
$D^L/Dt$ as
\begin{equation}\label{u-Xi-rel}
\mathbf{u}^\xi(\mathbf{x},t)
=
\Big(\frac{\partial}{\partial t}
+
\bar{\mathbf{u}}^L\cdot\nabla\Big)\,\Xi(\mathbf{x},t)
\equiv
\frac{D^L}{Dt}\,\Xi\,(\mathbf{x},t)
\,.
\end{equation}
Likewise, for any fluid quantity $\chi$ one may define $\chi^\xi$ as
the composition of functions
\begin{eqnarray*}
\chi^\xi(\mathbf{x},t)
=
\chi(\mathbf{x}^\xi,t)
=
\chi(\Xi\,(\mathbf{x},t),t)
\,.
\end{eqnarray*}
Taking the mean material time derivative and using the definition
of $D^L/Dt$ in equation (\ref{u-Xi-rel}) yields the {\bf advective
derivative relation},
\begin{eqnarray}\label{advect-der-eqn}
\frac{D^L}{Dt}\chi^\xi
&=&
\Big(\frac{\partial\chi}{\partial t}\Big)^\xi
+
T\chi\cdot\frac{D^L}{Dt}\Xi\,(\mathbf{x},t)
\nonumber\\
&=&
\Big(\frac{\partial\chi}{\partial t}
+
T\chi\cdot\mathbf{u}\Big)^\xi
\equiv
\Big(\frac{D\chi}{Dt}\Big)^\xi
\,.
\end{eqnarray}
As in equation (\ref{LA-vee}) for the velocity, the {\bf Lagrangian
mean} $\bar\chi^L$ of any other fluid quantity $\chi$ is defined as
\begin{eqnarray*}
\bar\chi^L(\mathbf{x},t)
\equiv
\overline{\chi^\xi(\mathbf{x},t)}
=
\overline{\chi(\mathbf{x}^\xi,t)}
=
\overline{\chi\big(\,\Xi\,(t)\cdot\mathbf{x},t\big)}
=
\overline{\chi\big(\,g(t)\cdot\mathbf{x}_0,t\big)}
\,.
\end{eqnarray*}
Taking the Lagrangian mean of equation (\ref{advect-der-eqn}) and once
again  using its projection property yields
\begin{eqnarray*}
\dot{\bar\chi}^L
=
\frac{D^L}{Dt}\bar\chi^L
=
\overline{\Big(\frac{D\chi}{Dt}\Big)}^L
=
\bar{\dot{\chi}}^L
\,,\quad\hbox{so that}\quad
\frac{D^L}{Dt}\chi^\ell
=
\Big(\frac{D\chi}{Dt}\Big)^\ell
\,,
\end{eqnarray*}
where $\chi^\ell=\chi^\xi-\bar\chi^L$ is the Lagrangian disturbance of
$\chi$ satisfying $\overline{\chi^\ell}=0$. 

\begin{rem}
The Lagrangian mean commutes with the material derivative.
Hence, the advective derivative relation (\ref{advect-der-eqn})
decomposes additively, as
\begin{equation}\label{advect-der-eqn-decomp}
\frac{D^L}{Dt}\Big(\chi^L+\chi^\ell\Big)
=
\overline{\Big(\frac{D\chi}{Dt}\Big)}^L
+
\Big(\frac{D\chi}{Dt}\Big)^\ell
\,.
\end{equation}

\end{rem}


\subsection{Mean advected quantities and their transformations} 


Advective transport by $g(t)$ and $\tilde{g}(t)$ is defined by
\begin{eqnarray*}
a(\mathbf{x}^\xi,t)
=
a_0\cdot g^{-1}(t)
\quad\hbox{and}\quad
\tilde{a}(\mathbf{x},t)
=
a_0\cdot \tilde{g}^{-1}(t)
\,,
\end{eqnarray*}
where $a_0=a(\mathbf{x}_0,0)=\tilde{a}(\mathbf{x}_0,0)$,
with $a,\tilde{a}\in V^*$ and the factorization $g(t)=\Xi\,(t)\cdot
\tilde{g}(t)$ implies 
\[
\tilde{a}(\mathbf{x},t)=a\cdot\,\Xi\,(\mathbf{x},t)
\,.
\]
Note that the right side of this equation is potentially rapidly
varying, but the left side is a mean advected quantity. 

Since $a$ and $\tilde{a}$ refer to the same
initial conditions, $a_0$, one finds
\begin{equation}\label{eff-def}
a_0\cdot \tilde{g}^{-1}(t)
=
\tilde{a}(\mathbf{x},t)
=
a\cdot\,\Xi\,(\mathbf{x},t)
=
a(\mathbf{x}^\xi,t)
\equiv
{\cal F}(\mathbf{x},t)\cdot a^\xi(\mathbf{x},t)
\,,
\end{equation}
where ${\cal F}(\mathbf{x},t)$ is the {\bf tensor transformation factor}
of $a$ under the change of variables
$\Xi\,:\mathbf{x}\to\mathbf{x}^\xi$. For example, one
computes formula (\ref{eff-def}) for an advected density as
\begin{equation}\label{advect-density}
\Big(D(\mathbf{x}_0)d^3x_0\Big)\cdot\tilde{g}^{-1}(t)
=
D^\xi(\mathbf{x},t) \det(T\Xi)\,d^3x
=
\tilde{D}(\mathbf{x},t)\,d^3x
\,.
\end{equation}
Thus, for an {\bf advected density}, $D$, 
\begin{equation}
D^\xi\det(T\Xi)(\mathbf{x},t) 
=
\tilde{D}(\mathbf{x},t)
\,,\quad 
{\cal F}(\mathbf{x},t) = \det(T\Xi)
\,,\quad
\frac{\partial}{\partial t}\tilde{D}
=
-\,
{\rm div}\,(\tilde{D}\tilde{\mathbf{u}})
\,.
\nonumber
\end{equation}
For an {\bf advected scalar function}, $s$, 
\begin{equation}
s^\xi(\mathbf{x},t) 
=
\tilde{s}(\mathbf{x},t)
=
\bar{s}^L(\mathbf{x},t)
\,,\quad
{\cal F}=1
\,,\quad
\frac{\partial}{\partial t}\tilde{s}
=
-\,
\tilde{\mathbf{u}}\cdot \nabla \tilde{s}
\,.
\nonumber
\end{equation}
For an {\bf advected vector field}, $\mathbf{B}$, 
\begin{equation}
K^i_j\,{B}^{\xi\,j}(\mathbf{x},t) 
=
\tilde{B}^i(\mathbf{x},t)
\,,\quad\hbox{and}\quad 
K^i_j = \det(T\Xi\,)\,(T\Xi^{\,-1})^i_j
\,.
\nonumber
\end{equation}
Thus, in the case of an advected vector field, one has
\begin{equation}
\mathsf{K}\cdot\mathbf{B}^\xi(\mathbf{x},t)
=
\tilde{\mathbf{B}}(\mathbf{x},t)
\,,\quad\hbox{with}\quad 
{\cal F}=\mathsf{K}\equiv\det(T\Xi\,)\,T\Xi^{\,-1}
\,,
\nonumber
\end{equation}
and an advection relation (e.g., a frozen-in magnetic field)
given by
\begin{equation}
\frac{\partial}{\partial t}\tilde{\mathbf{B}}
=
-\,
\tilde{\mathbf{u}}\cdot \nabla \tilde{\mathbf{B}}
+
\tilde{\mathbf{B}}\cdot \nabla \tilde{\mathbf{u}}
\,.
\nonumber
\end{equation}

\noindent
Finally, for an {\bf advected symmetric tensor} $S$ one finds
\begin{equation}
(T\Xi^{\,T}\cdot S^\xi\cdot T\Xi\,)_{ij}
=
\tilde{S}_{ij}
\,,
\nonumber
\end{equation}
whose advection relation is obtained as in the other cases.  

In each case, the corresponding transformation factor ${\cal F}$ appears
in a    {\bf variational relation} for an advected quantity, expressed
via equation (\ref{eff-def}) as
\begin{equation}\label{var-a-rel}
\delta{a}^\xi
=
\delta\,({\cal F}^{-1}\cdot\tilde{a})
=
{\cal F}^{-1}\cdot\delta\tilde{a}
+
(\delta{\cal F}^{-1})\cdot\tilde{a}
\,.
\end{equation}
This formula will be instrumental in establishing the main result.


\subsection{Lagrangian Averaged Euler-Poincar\'e Theorem} 


Let the assumptions hold as listed previously for the EP
Theorem \ref{rarl} and assume the GLM factorization  
$g(t)=\Xi\,(t)\cdot\tilde{g}(t)$ with $\bar{g}(t)
=
\overline{\Xi\,(t)\cdot \tilde{g}(t)}$. Then,

\begin{theorem} [LAEP Theorem]\label{LAEP-Thm}


The following are equivalent:
\begin{enumerate}

\item [$\overline{\bf i}$ ] The averaged Hamilton's principle holds
\begin{equation} \label{Avg-Ham-Princ}
\delta \int _{t_1} ^{t_2} 
\overline{
L_{a_0}(g(t), \dot{g} (t))
}
\, dt 
=
0
\end{equation}
for variations $\delta g(t)$
of $ g (t) $ vanishing at the endpoints.

\item [$\overline{\bf ii}$  ] The mean Euler--Lagrange
equations for $\bar{L}_{a_0}$ are satisfied on $\tilde{G}$,
\begin{equation} \label{Avg-EL-eqns}
\overline{
\frac{\delta L_{a_0}}{\delta {g}}\cdot T\Xi\,
}
-
\overline{
\frac{d}{dt}
\frac{\delta L_{a_0}}{\delta \dot{g}}
\cdot T\Xi
}
=
0
\end{equation}
%

\item [$\overline{\bf iii}$ ]  The averaged constrained variational
principle
\begin{equation} \label{Avg-var-princ}
\delta \int _{t_1} ^{t_2}  
\overline{
\ell\,\big(u^\xi(t), a^\xi(t)\big)
} dt = 0
\end{equation}
holds on $\tilde{\mathfrak{g}} \times \tilde{V} ^\ast $, using
variational relations of the form
\begin{eqnarray*} \label{avg-variationsright}
\delta u^\xi
\!\!&=&\!\!
T\Xi\,\cdot \! \Big(
\frac{ \partial \tilde{\eta} }{\partial t } 
+ {\rm ad}_{\tilde{u}}\,\tilde{\eta}
\Big)
+
\delta\,\Xi\hbox{ terms}
\,,
\\
\delta a^\xi
\!\!&=&\!\!
{\cal F}^{-1} \! \cdot\delta\tilde{a}
+
\delta\,\Xi\hbox{ terms, and }
\delta\tilde{a}
=
  -\,\tilde{a}\,\tilde{\eta} ,
\end{eqnarray*}
where 
$\tilde{\eta}(t) = \delta \tilde{g}\,\tilde{g}^{-1}
\in \tilde{\mathfrak{g}}$ vanishes at
the endpoints.

\item [$\overline{\bf iv}$] The Lagrangian averaged
Euler--Poincar\'{e} (LAEP) equations hold on
$\tilde{\mathfrak{g}} \times \tilde{V}^\ast$
\begin{equation} \label{LAEP-eqns-right}
\frac{ \partial}{\partial t} 
\overline{\Big(
\frac{\delta \ell}{\delta u^\xi}
\cdot T\Xi\Big)}
= 
-
 {\rm ad}_{\tilde{u}}^{\ast} 
\overline{\Big(
\frac{\delta \ell}{\delta u^\xi}
\cdot T\Xi\Big)}
+ 
\overline{\Big(
\frac{\delta \ell}{\delta a^\xi} 
\cdot {\cal F}^{-1}\Big)}
\diamond \tilde{a}
\,.
\end{equation}
\end{enumerate}

\end{theorem}


\begin{corollary} [LA Kelvin-Noether Circulation
Theorem]\label{LA-Kel-Thm}
\begin{equation}
\frac{d}{dt}
\oint_{c(\tilde{u})}
\frac{1}{\tilde{D}}\
\overline{\Big(
\frac{\delta \ell}{\delta u^\xi}
\cdot T\Xi\Big)}
=
\oint_{c(\tilde{u})}
\frac{1}{\tilde{D}}\
\overline{\Big(
\frac{\delta \ell}{\delta a^\xi} 
\cdot {\cal F}^{-1}\Big)}
\diamond \tilde{a}
\,,
\nonumber
\end{equation}
for any closed curve $c(\tilde{u})$ that moves with the fluid.

\end{corollary}
\bigskip

\noindent
{\bf Proof.}
Via the equivalence of ad$^*$ and Lie derivative for a one-form
density, the (LAEP) equation implies
\begin{equation}
\frac{d}{dt}
\oint_{c(\tilde{u})}
\frac{1}{\tilde{D}}\
\overline{\Big(
\frac{\delta \ell}{\delta u^\xi}
\cdot T\Xi\Big)}
=
\oint_{c(\tilde{u})}\!\!
\Big(
\frac{ \partial}{\partial t}
+
\pounds_u
\Big)
\frac{1}{\tilde{D}}\
\overline{\Big(
\frac{\delta \ell}{\delta u^\xi}
\cdot T\Xi\Big)}
\,,
\nonumber
\end{equation}
for any closed curve $c(\tilde{u})$ that moves with the fluid.
\par\hfill QED\smallskip

\subsection{Proof of the LAEP Theorem}



The equivalence of $\overline{\bf i}$ and $\overline{\bf ii}$ holds
for any configuration manifold and so, in particular, it holds again in
this case. To compute the averaged Euler-Lagrange equation
(\ref{Avg-EL-eqns}), we use the following {\bf variational relation}
obtained from the composition of maps $g(t)=\Xi\,(t)\cdot
\tilde{g}(t)$, cf. the velocity relation (\ref{vee-rel}),
\begin{equation}\label{var-gee-rel}
\delta{g}(t)
=
\delta{\Xi}\,(t)\cdot\tilde{g}(t)
+
T\Xi\,(t)\cdot \delta\tilde{g}(t)
\,.
\end{equation}
Hence, we find
\begin{eqnarray*} \label{Avg-Ham-Princ-proof}
0 
&=& 
\delta \int _{t_1} ^{t_2} 
\overline{
L_{a_0}(g(t), \dot{g} (t))
}
\, dt 
\\
&=&
\int _{t_1} ^{t_2} 
\bigg(\
\overline{
\frac{\delta L_{a_0}}{\delta {g}}\cdot \delta {g} 
}
+
\overline{
\frac{\delta L_{a_0}}{\delta \dot{g}}\cdot \delta \dot{g} 
}\
\bigg)
\,dt
\\
&=&
\int _{t_1} ^{t_2} 
\bigg(\
\overline{
\frac{\delta L_{a_0}}{\delta {g}}\cdot T\Xi\,
}
-
\overline{
\frac{d}{dt}
\frac{\delta L_{a_0}}{\delta \dot{g}}
\cdot T\Xi}\
\bigg)\cdot\delta\tilde{g}\,dt
\,.
\end{eqnarray*}
This yields the mean Euler-Lagrange equations (\ref{Avg-EL-eqns}).
Here we have dropped $\delta\,\Xi-$terms, because they do not figure
in the variational principle for Lagrangian mean fluid dynamics at
this level of description.

The following string of equalities shows that $\overline{\bf iii}$ is
equivalent to $\overline{\bf iv}$.
\begin{eqnarray}\label{avg-EPderivation}
0 
&=&
\delta \int_{t_1}^{t_2}
   \overline{\ell(u^\xi, a^\xi)}\, dt 
=
\int_{t_1}^{t_2}
\Big\langle
\overline{
\frac{\delta \ell}{\delta u^\xi}\,,\,\delta u^\xi 
}
\Big\rangle
+
\Big\langle
\overline{
\frac{\delta \ell}{\delta a^\xi}\,,\,\delta a^\xi 
}
\Big\rangle
\,dt 
\nonumber \\
  &=& 
\int_{t_1}^{t_2}
\Big\langle
\overline{
\frac{\delta \ell}{\delta u^\xi}\,,\,
T\Xi\,\cdot \delta(\dot{\tilde{g}}\tilde{g}^{-1})
}
\Big\rangle
+
\Big\langle
\overline{
\frac{\delta \ell}{\delta a^\xi}\,,\,
{\cal F}^{-1} \! \cdot\delta\tilde{a}
}
\Big\rangle
\,dt 
\nonumber \\
  &=& 
 \int_{t_1}^{t_2}
  \left\langle\
\overline{
\frac{\delta \ell}{\delta u^\xi}
\cdot T\Xi}
  \,,\, 
  \frac{\partial\tilde{\eta} }{\partial t}
    +{\rm ad}_{\tilde{u}}\,\tilde{\eta} 
 \right\rangle
    -
\left\langle\
\overline{
\frac{\delta \ell}{\delta a^\xi} 
\cdot {\cal F}^{-1}}
\,,\, \pounds_{\tilde{\eta}}\, \tilde{a}
\right\rangle
\,dt
\nonumber \\
   &&\hspace{-12mm}=-\int_{t_1}^{t_2}
 \left\langle
\frac{ \partial}{\partial t} 
\overline{\Big(
\frac{\delta \ell}{\delta u^\xi}
\cdot T\Xi\Big)}
+
 {\rm ad}_{\tilde{u}}^{\ast} 
\overline{\Big(
\frac{\delta \ell}{\delta u^\xi}
\cdot T\Xi\Big)}
- 
\overline{\Big(
\frac{\delta \ell}{\delta a^\xi} 
\cdot {\cal F}^{-1}\Big)}
\diamond \tilde{a}
\,,\,
\tilde{\eta}
\right\rangle
dt
\,.
\nonumber
\end{eqnarray}
In the second line, we again dropped the $\delta\,\Xi-$terms and we
substituted the following {\bf variational relations} obtained from
equations (\ref{u-xi-def}) and (\ref{var-a-rel}), 
\begin{eqnarray} \label{avg-variations-proof1}
\delta u^\xi
&=& 
T\Xi\,\cdot \delta(\dot{\tilde{g}}\tilde{g}^{-1})
+
\delta\,\Xi\hbox{ terms}
\\
&=&
T\Xi\,\cdot \! \Big(
\frac{ \partial \tilde{\eta} }{\partial t } 
+ {\rm ad}_{\tilde{u}}\,\tilde{\eta}
\Big)
+
\delta\,\Xi\hbox{ terms}
\,,
\label{avg-variations-proof2}
\\
\delta a^\xi
&=&
{\cal F}^{-1} \! \cdot\delta\tilde{a}
+
\delta\,\Xi\hbox{ terms}
\quad\hbox{and}\quad
\delta\tilde{a}
=
  -\tilde{a}\,\tilde{\eta} 
\,.\label{avg-variations-proof3}
\end{eqnarray}

Finally we show that $\overline{\bf i}$ and $\overline{\bf iii}$ are
equivalent. First note that the $G$--invariance of $L:TG
\times V^\ast \rightarrow \mathbb{R}$ and the definition of 
$a(t) = a_0g(t)^{-1}$ imply that the
integrands in (\ref{Avg-Ham-Princ}) and
(\ref{Avg-var-princ}) are equal, both before and after averaging.
Moreover, all variations $\delta g(t) \in TG$ of $g(t)$ with fixed
endpoints induce and are induced by variations $\delta u(t) \in
\mathfrak{g}$ of $u(t)$ of the form $\delta u = \partial\eta
/\partial t + {\rm ad}_u\,\eta $ with $\eta(t) \in \mathfrak{g}$
vanishing at the endpoints. The relation between $\delta g(t)$ and
$\eta(t)$ is given by $\eta(t) = \delta g(t)g(t)^{-1}$. The
corresponding statements are also true for the tilde-variables in
the variational relations (\ref{var-gee-rel}) and
(\ref{avg-variations-proof1}) -- (\ref{avg-variations-proof3}) that
are used in the calculation of the other equivalences.
\par\hfill QED\smallskip

\begin{rem}[Lagrangian Average Conservation Laws/Balances]

From the viewpoint of the LAEP theorem, the Kelvin circulation
theorem and its associated conservation of potential vorticity for LA
flows both emerge because reduction of Hamilton's principle by its
relabeling symmetries in passing from the material to the spatial
picture of continuum mechanics is compatible with Lagrangian
averaging, which takes place at {\it fixed} fluid labels.

LA also preserves the kinematic symmetries of Hamilton's principle,
so, conservation, or balance, laws for momentum and energy for the LA
dynamics are also guaranteed by Noether's theorem for the averaged
variational principle, according to its transformations under space and
time translations.

\end{rem}



\section{Application of the LAEP theorem to incompressible fluids}
\label{Applic-incomp-sec}


\subsection{Euler's equation for an incompressible fluid}

For an incompressible fluid, the EP theorem \ref{rarl} yields Euler's
equations as
\begin{equation} \label{euler-ep}
\frac{ \partial}{\partial t} \frac{\delta \ell}{\delta u} 
= 
-
 {\rm ad}_{u}^{\ast} \frac{ \delta \ell }{ \delta u}
+ 
\frac{\delta \ell}{\delta D} \diamond D
\,,
\end{equation}
for the reduced Lagrangian 
\begin{equation} \label{reduced-euler-lag}
\ell
= 
\int 
\frac{1}{2}D|u|^2 - p\,(D-1)
\,d^3x
\,.
\end{equation}
Here the pressure $p$ is a Lagrange multiplier that imposes
incompressibility. The variational derivatives of this Lagrangian are
given by
\begin{equation} \label{reduced-euler-lag-var}
\delta\ell
= 
\int 
Du\cdot\delta u 
+
\Big(\frac{1}{2}|u|^2 - p \Big)\delta D
- (D-1)\delta p
\
d^3x
\,.
\end{equation}
The expected Euler equation for incompressible fluids is found upon
setting $D=1$ in equation (\ref{euler-ep}) as
\begin{equation} \label{euler-eqn}
\frac{ \partial}{\partial t} u
+
u\cdot \nabla u 
+
\nabla p
= 
0\,.
\end{equation}
The auxiliary advection relation for the mass density $D$ is the
continuity equation 
\begin{equation}
\frac{ \partial D}{\partial t}
=
-\,{\rm div}(Du)
\,,
\end{equation}
which, as usual, ensures incompressibility via the constraint $D=1$.


\subsection{The Lagrangian averaged Euler (LAE) equations}

The Lagrangian averaged Euler (LAE) equations are derived from the
LAEP theorem \ref{LAEP-Thm} as follows. The corresponding averaged
Lagrangian in the {\it material description} is given by
\begin{equation} \label{avg-material-lag}
\bar{L}
= 
\int 
D_0\,d^3x_0\
\Big[\
\frac{1}{2}\,\overline{\,|\dot{\mathbf{x}}^\xi|^2}
+
\overline{p^\xi\,
\Big(\det\frac{\partial \mathbf{x}^\xi}{\partial \mathbf{x}_0}-1\Big)
}\Big]
\,.
\end{equation}
Therefore, the reduced averaged Lagrangian in the
{\it spatial picture} becomes 
\begin{equation} \label{reduced-avg-spatial-lag}
\bar\ell
= 
\int d^3x\
\Big[\
\frac{1}{2}\,\tilde{D}\,
\overline{ \,|\mathbf{u}^\xi|^2 }
+
\overline{p^\xi\,
\Big(\det T\Xi\,-\tilde{D}\Big)
}\
\Big]
\,,
\end{equation}
where we have used equation (\ref{advect-density}) in the change of
variables. The necessary variations of this Lagrangian are
given by (dropping the $\delta\,\Xi-$ terms)
\begin{eqnarray} \label{avg-euler-lag-var}
\delta\bar\ell
&=& 
\int d^3x\
\Big[\
\tilde{D}\,
\overline{ \,\mathbf{u}^\xi\cdot T\Xi\,}\cdot\delta\tilde{\mathbf{u}}
+
\Big(\,\frac{1}{2}\,\overline{\, |\mathbf{u}^\xi|^2} - \bar{p}^L \Big)
\delta \tilde{D}
\\
&&\hspace{20mm}+\
\delta \bar{p}^L \Big(\
\overline{ \,\det T\Xi\,}
-
\tilde{D}
\Big)
+
\overline{\delta p^\ell\,\det T\Xi\,}\
\Big]
\,.
\nonumber
\end{eqnarray}
Here we substituted the pressure decomposition
$p^\xi=\bar{p}^L+p^\ell$ with $\overline{p^\xi}=\bar{p}^L$ and used
the projection property of the Lagrangian average. Thus, the pressure
constraint implies that the mean advected density is related to the
mean fluid trajectory by 
\begin{equation} 
\tilde{D} = \overline{ \,\det T\Xi\,}
\,.
\nonumber
\end{equation}
Consequently, the LAE fluid velocity in general has a
divergence, 
\begin{equation} 
{\rm div}\,\tilde{\mathbf{u}}\ne0
\,,
\nonumber
\end{equation}
as was first noticed in Andrews {and} McIntyre [1978a].
The Lagrangian disturbance of the pressure $p^\ell$ imposes the
constraint
\begin{equation} 
\overline{\delta p^\ell\,\det T\Xi\,} = 0
\,.
\nonumber
\end{equation}
This constraint also arises in the self-consistent theory of
wave-mean flow interaction dynamics in Gjaja {and} Holm [1996].
It is irrelevant here, though, because we are not considering
self-consistent fluctuation dynamics. (The self-consistent theory
arises from the $\delta\,\Xi-$terms that we dropped here.)

The LAE equation may now be written in LAEP form
(\ref{LAEP-eqns-right}) in components as 
\begin{equation} \label{LAEP-eqns}
\frac{ \partial}{\partial t} \tilde{v}_i
+
\tilde{u}^j\frac{ \partial}{\partial x^j}\tilde{v}_i
+
\tilde{v}_j\frac{ \partial}{\partial x^i}\tilde{u}^j
+
\frac{ \partial}{\partial x^i}\tilde{\pi}
= 
0
\,,\end{equation}
\begin{equation} \label{LAEP-defs}
\tilde{v}_i
=
\frac{1}{\tilde{D}}
\frac{\delta \ell}{\delta \tilde{u}^i}
= 
\overline{u^\xi_j(T\Xi)^j_i}
\,,\quad
\tilde{\pi}=-\,\frac{\delta \ell}{\delta \tilde{D}}
=
-\,\frac{1}{2}\,\overline{\, |\mathbf{u}^\xi|^2} + \bar{p}^L
\,,
\end{equation}
and the advected mean mass density $\tilde{D}$ satisfies the
corresponding mean continuity equation 
\begin{equation}\label{LAEP-continuity-eqn}
\frac{ \partial\tilde{D}}{\partial t}
=
-\,{\rm div}(\tilde{D} \tilde{\mathbf{u}})
\,.
\end{equation}
When $T\Xi = Id + \nabla \xi$, one finds 
\begin{equation} \label{LAEP-GLM}
\tilde{\mathbf{v}}
= 
\overline{\mathbf{u}^\xi}
+
\overline{\frac{D^L}{Dt}\xi_j\nabla\xi^j}
\equiv
\bar{\mathbf{u}}^L - \bar{\mathbf{p}}
\,.
\end{equation}
The term $\bar{\mathbf{p}}$ is called the pseudomomentum in 
Andrews {and} McIntyre [1978a]. See, e.g., Holm [2001] for a
recent discussion and more details.

\begin{rem}[Momentum balance]
The EP theory of Holm, Marsden {and} Ratiu [1998a] implies
momentum balance in this case in the form,
\begin{equation} \label{LAEP-mom-conserv}
\frac{ \partial}{\partial t} (\tilde{D}\tilde{v}_i)
+
\frac{ \partial}{\partial x^j}\Big(
\tilde{D}\tilde{v}_i\tilde{u}^j
+
\bar{p}^L\delta^j_i\Big)
= 
\frac{\tilde{D}}{2}
\frac{ \partial\overline{\,|\mathbf{u}^\xi|^2}}{\partial
x^i}\,\bigg|_{exp}\,,
\end{equation}
where subscript $exp$ refers to the {\bf explicit} spatial
dependence arising from the $\Xi-$terms in
$\overline{\,|\mathbf{u}^\xi|^2}
=
\overline{\,|D^L\,\Xi\,/Dt\,|^2}$ obtained from equation
(\ref{u-xi-def}).
\end{rem}


\subsection{Recent progress toward closure}

Of course, the LAE equations (\ref{LAEP-eqns}) --
(\ref{LAEP-continuity-eqn}) are not yet closed. As indicated in
their momentum balance relation (\ref{LAEP-mom-conserv}), they depend on
the unknown Lagrangian statistical properties appearing as the
$\Xi-$terms in the definitions of $\tilde{\mathbf{v}}$ and
$\tilde{\pi}$. Until these properties are modeled or prescribed, the
LAE equations are incomplete.

Progress in formulating and analyzing a closed system of fluid equations
related to the LAE equations has recently been made in the EP context.
These closed model LAE equations were first obtained in Holm, Marsden
{and} Ratiu [1998a,b].  For more discussion of this type of
equation and its recent developments as a turbulence model, see papers
by Chen {\it et al} [1998, 1999a,b,c], Shkoller [1998], 
Foias {\it et al} [1999],[2001] and Marsden, Ratiu {and} Shkoller [2001]
and Marsden {and} Shkoller [2001].  An earlier self-consistent variant of
the LAE closure was also introduced in Gjaja {and} Holm [1996]. This was
further developed in Holm [1999,2001]. 

\begin{rem}
[Transport structure]
Although the LAE equations are not yet closed, their transport
structure may still be discussed because they are derived in the LAEP
context, which preserves the transport structure. Thus, as in equations
(\ref{Kelvin-circ-eqn}) and (\ref{Helicity-eqn}) we have 
\begin{equation} \label{LAE-kelvin-them}
\frac{d}{dt}\oint_{c(\tilde{\mathbf{u}})}
\tilde{\mathbf{v}}\cdot d\mathbf{x}
= 
0
\,,
\quad(\hbox{LAE Kelvin theorem})
\,,
\end{equation}
and
\begin{equation} \label{LAE-helicity-cons}
\frac{d}{dt}\int
\tilde{\mathbf{v}}\cdot {\rm curl}\,\tilde{\mathbf{v}}
\,d^3x
= 
0
\,
\quad(\hbox{LAE Helicity conservation})
\,.
\end{equation}
Of course, the LAEP approach is versatile enough to derive LA equations
for compressible fluid motion, as well. This was already shown in the
GLM theory of Andrews {and} McIntyre [1978a]. For brevity, we
only remark that the LAEP approach also preserves magnetic helicity and
cross-helicity conservation when applied to magnetohydrodynamics (MHD). 
\end{rem}



\section{Acknowledgements}

I am grateful for stimulating discussions of this topic with P.
Constantin, G. Eyink, U. Frisch, J. Marsden, M. E. McIntyre, I. Mezic,
S. Shkoller and A. Weinstein. Some of these discussions took place at
Cambridge University while the author was a visiting professor at the
Isaac Newton Institute for Mathematical Science. This work was
supported by the U.S. Department of Energy under contracts
W-7405-ENG-36 and the Applied Mathematical Sciences Program KC-07-01-01.


\begin{description}

\item
Andrews, D G {and} McIntyre, M E 1978
An exact theory of nonlinear waves on a Lagrangian-mean flow.
{\it J. Fluid Mech.} {\bf 89} 609--646.

\item
Arnold, V I 1966
Sur la g\'{e}ometrie differentielle des groupes de Lie de dimension
infinie et ses applications \`{a} l'hydrodynamique des fluides
parfaits.
{\it Ann. Inst. Fourier (Grenoble)} {\bf 16} 319--361.

\item
Chen, S Y, Foias, C, Holm, D D, Olson, E J, Titi, E S {and}
Wynne, S 1998
The Camassa-Holm equations as a closure model for
turbulent channel and pipe flows.
{\it Phys. Rev. Lett.} {\bf 81} 5338-5341.

\item
Chen, S Y, Foias, C, Holm, D D, Olson, E J, Titi, E S {and}
Wynne, S 1999
The Camassa-Holm equations and turbulence in pipes and channels.
{\it Physica D} {\bf133} 49-65.

\item
Chen, S Y, Foias, C, Holm, D D, Olson, E J, Titi, E S {and}
Wynne, S (1999)
A connection between the Camassa-Holm equations and turbulence
in pipes and channels. 
{\it Phys. Fluids} {\bf11} 2343-2353.

\item
Chen, S Y, Holm, D D, Margolin, L G {and} Zhang, R (1999)
Direct numerical simulations of the Navier-Stokes alpha model.
{\it Physica D} {\bf133} 66-83.

\item
Foias, C, Holm, D D {and} Titi, E S (1999)
The Three Dimensional Viscous Camassa--Holm Equations,
and Their Relation to the Navier--Stokes Equations and
Turbulence Theory.
{\it J. Diff. Eq.} at press.

\item
Foias, C, Holm, D D {and} Titi, E S (2001)
The Navier-Stokes-alpha model of fluid turbulence.
{\it Physica D} at press.

\item
Gjaja, I {and} Holm, D D (1996)
Self-consistent wave-mean flow interaction
dynamics and its Hamiltonian formulation for a rotating
stratified incompressible fluid.
{\it Physica D} {\bf 98} 343-378.

\item
Holm, D D (1999)
Fluctuation effects on 3D Lagrangian mean
and Eulerian mean fluid motion.
{\it Physica D} {\bf133} 215-269.

\item
Holm, D D (2001)
Averaged Lagrangians and the mean dynamical effects 
of fluctuations in continuum mechanics. 
{\it Physica D} at press.

\item
Holm, D D, Marsden, J E {and} Ratiu, T S (1998)
The Euler--Poincar\'{e} equations and semidirect products
with applications to continuum theories.
{\it Adv. in Math.} {\bf 137} 1-81.

\item
Holm, D D, Marsden, J E {and} Ratiu, T S (1998)
Euler--Poincar\'e models of ideal fluids
with nonlinear dispersion,
{\it Phys. Rev. Lett.} {\bf 80} 4173-4177.

\item
Holm, D D, Marsden, J E, Ratiu, T S {and}  Weinstein, A (1985)
Nonlinear stability of fluid and plasma equilibria.
{\it Physics Reports} {\bf 123} 1--116.

\item
Marsden, J E {and} Ratiu, T S (1999)
{\it Introduction to Mechanics and Symmetry} 
Springer: New York, 2nd Edition.

\item
Marsden, J E {and} Shkoller, S (2001)
The anisotropic averaged Euler equations.
{\it J. Rat. Mech. Anal.} at press.

\item
Marsden, J E, Ratiu, T S {and} S Shkoller (2001) 
The geometry and analysis of the averaged
Euler equations and a new diffeomorphism group. 
{\it Geom. Funct. Anal.} at press.

\item
Shkoller, S (1998)
Geometry and curvature of diffeomorphism groups
with $H_1$ metric and mean hydrodynamics.
{\it J. Funct. Anal.} {\bf160} 337--365.

\end{description}

\end{document}